\documentclass[twocolumn,prl,floatfix,superscriptaddress]{revtex4}
\usepackage{color}
\usepackage{epsfig}
\usepackage{graphicx}
\usepackage{amsmath}
\usepackage[applemac]{inputenc}
\usepackage{ae, aeguill}

\newcommand{\angstrom}{\mbox{\normalfont\AA}}

\newcommand {\be}  {
\begin{equation}
}
\newcommand {\ee}  {
\end{equation}
}
\newcommand {\bea} {
\begin{eqnarray}
}
\newcommand {\eea} {
\end{eqnarray}
}
\newcommand{\RNum}[1]{\uppercase\expandafter{\romannumeral #1\relax}}

\begin{document}

\title{Effects of aluminum on hydrogen solubility and diffusion in deformed Fe-Mn alloys}

\author{C. H\"uter}
\affiliation{Institute for Energy and Climate Research, Forschungszentrum J\"ulich GmbH, 52425 J\"ulich, Germany}
\affiliation{Computational Materials Design Department, Max-Planck Institut f\"ur Eisenforschung, 40237 D\"usseldorf, Germany}
\author{S. Dang}
\affiliation{Institute for Energy and Climate Research, Forschungszentrum J\"ulich GmbH, 52425 J\"ulich, Germany}
\author{X. Zhang}
\author{A. Glensk}
\affiliation{Computational Materials Design Department, Max-Planck Institut f\"ur Eisenforschung, 40237 D\"usseldorf, Germany}
\author{R. Spatschek}
\affiliation{Institute for Energy and Climate Research, Forschungszentrum J\"ulich GmbH, 52425 J\"ulich, Germany}
\affiliation{Computational Materials Design Department, Max-Planck Institut f\"ur Eisenforschung, 40237 D\"usseldorf, Germany}
        
\begin{abstract}
We discuss hydrogen diffusion and solubility in aluminum alloyed Fe-Mn alloys. 
The systems of interest are subjected to  tetragonal and isotropic deformations. 
Based on {\it{ab initio}} modelling, we calculate solution energies, then employ Oriani's theory which reflects the influence of Al alloying via trap site diffusion. 
This local equilibrium model is complemented by qualitative considerations of Einstein diffusion. 
Therefore, we apply the climbing image nudged elastic band method to compute the minimum energy paths and energy barriers for hydrogen diffusion. 
Both for diffusivity and solubility of hydrogen, we find that the influence of the substitutional Al atom has both local chemical and nonlocal volumetric contributions. 
\end{abstract}          

\date{\today}

\maketitle

\section{\RNum{1}. Introduction}
High strength structural steels are a material class key to further developments in automotive light-weight construction  \cite{koyama2014hydrogen,scott2006development,de2011high}. 
At Mn weight fractions between 15-25 percent, high manganese steels offer a combination of high strength and desirable plasticity characteristics. The stable austenitic structure offers high cost efficiency \cite{barbier2009analysis,koyama2013twip,gutierrez2011dislocation,steinmetz2013revealing,koyama2011work}, and investigations on Fe-Mn systems revealed the relation of high work hardening rates and ductility to the austenitic microstructure \cite{allain2004physical,frommeyer2003supra}. In addition, high manganese steels exhibit a lower susceptibility to hydrogen embrittlement \cite{grajcar2010corrosion} than ferritic and martensitic steels \cite{frohmberg1954delayed,perng1987comparison}. 

The systematic investigation of hydrogen embrittlement in high manganese steels is challenged by the microstructural, interfacial and chemical complexity of these materials. The scientific focus has been on intergranular failure modes and general heterophase interface failure \cite{park2012mechanism,koyama2013effects,koyama2012hydrogen,koyama2013hydrogen,koyama2012hydrogen2}.  Hydrogen induced degradation \cite{grajcar2012corrosion} is a major hindrance for the implementation of promising steel designs for automotive applications  \cite{grajcar2012third}. While the austinitic phase itself exhibits low diffusivity and high solubility of hydrogen, the transformation induced plasticity (TRIP) moderated martensitic transformation yields strong gradients of solubility and diffusivity due to the presence of ferritic phases.

An improved resistance to hydrogen induced delayed fracture can be realised adding temper softeners like Cr, Mo and V in martensitic steels. This highly desirable effect originates from the suppressed fracture by carbide precipitation, shifting the intergranular fracture to intragranular fracture \cite{hata2014cold}. In austenitic steels, the addition of Ni and Si is an established method to increase the ductility of hydrogen-loaded steels \cite{michler2008hydrogen,louthan1981environmental,san2008effects,gavriljuk2003diagnostic}, and the 
beneficial effect of Mn on hydrogen solubility has been investigated for instance in \cite{ismer2010ab, von-Appen:2014aa}. The successful 
alloying of high manganese steels with Al, which reduces the weakness against hydrogen embrittlement, suggests 
to study the effect of Al on the different aspects of hydrogen in Fe-Mn systems. As it is an integral part of production processes, 
we also study the influence of mechanical loading, in practice isotropic and tetragonal distortions, on solubility and mobility of hydrogen. 

The article is organized as follows:
In the second section we describe the methodical basis of our approach, including several aspects of the 
systematic approximations we make. 
The third section contains the results of our investigations, where we distinguish between stress-free, isotropically distorted and tetragonally distorted systems and discuss their influence on hydrogen solubility and diffusivity. 
Finally we conclude with a brief discussion of the presented results in the context of different material configurations and future perspectives.

\section{\RNum{2}. Methods}
The methodical basis of our investigation is electronic structure density functional theory (DFT). 
It allows to predict the quantities of interest for exactly defined crystal geometries and chemical compositions. 
\begin{figure}
\begin{center}
\includegraphics[trim=20.5cm 2.5cm 20.5cm 2.5cm, clip=true, width=6cm]{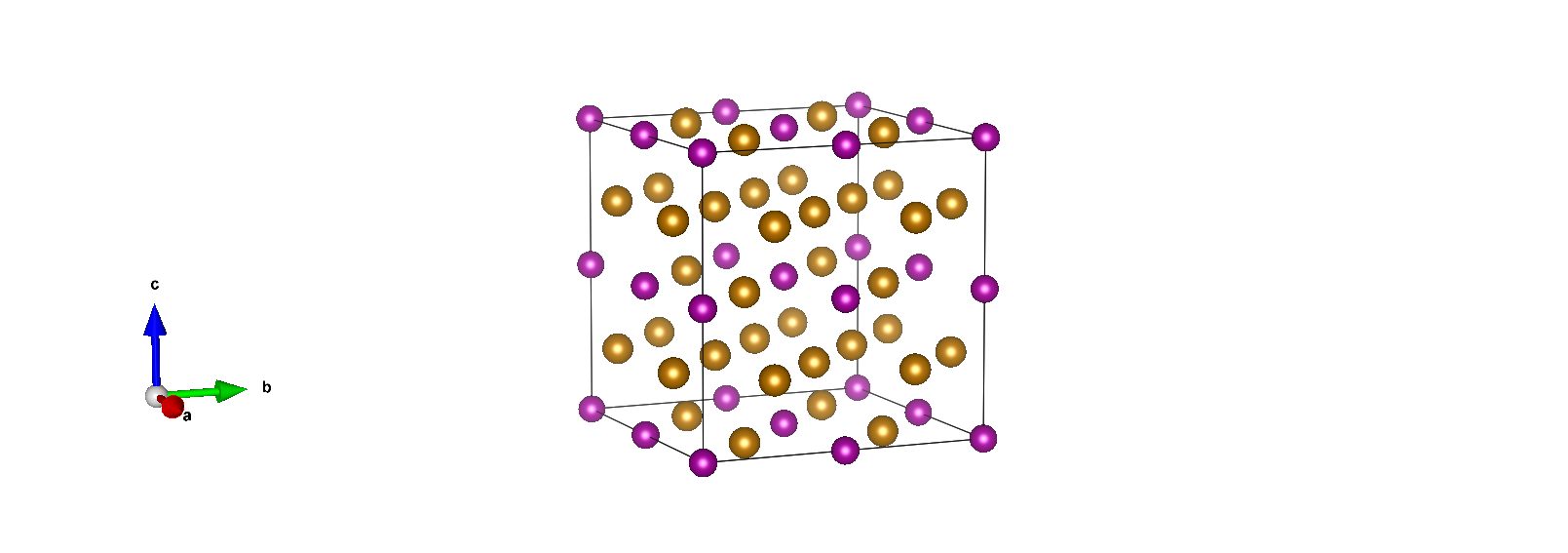}
\caption{The supercell of the austenitic Fe-Mn system, Fe atoms are shown in yellow and Mn atoms in purple. The distribution of the Mn atoms yields only two types of octahedral interstices  and one inequivalent site for the substitution Fe $\to$ Al which makes it an ideal reference for the study of the Fe$_{23}$Mn$_8$Al system.}
\label{fig10}
\end{center}
\end{figure}
The systems which we investigate are introduced in Figs.~\ref{fig10} and \ref{octaSites}.
They consist of  2$\times$2$\times$2 supercells of an austentitic Fe$_{24}$Mn$_8$ matrix and the resulting system when we substitute one Fe atom by an Al atom, Fe$_{23}$Mn$_8$Al. 
Hydrogen is introduced on interstitial sites.
We obtained our {\em ab initio} results from DFT calculations employing the 
Vienna ab initio simulation package (VASP) \cite{kresse1996software}. 
For the PAW pseudo-potentials \cite{blochl1994projector} 
we use the GGA-PBE exchange and correlation functional \cite{perdew1996generalized}. The convergence of DFT related parameters has been carefully tested.  We find that an 8$\times$8$\times$8 k-point Monkhorst-Pack \cite{monkhorst1976special} mesh for the 2$\times$2$\times$2 supercell and a plane wave energy cutoff of 600 eV are sufficient to ensure a convergence of the calculated formation energies to better than 10 meV. 
An electronic smearing of 0.1 eV was chosen within the Methfessel-Paxton scheme \cite{methfessel1989high}.
Further, subject-specific numerical details will be discussed at the appropriate places of this publication.

\section{\RNum{3}. Results}
\subsection{Hydrogen solubility}

The goal of this section is to determine the concentration of hydrogen as solid solution in the fcc Fe-Mn-Al system.
Two aspects are of primary interest here:
First, the local arrangement of the hydrogen atoms in the interstitial sites in different environments of iron, manganese and aluminum, to find a site preference for the hydrogen.
Second, the dependence of the solubility on mechanical deformation conditions, in particular for pure volume changes due to isotropic straining, and for volume-conserving tetragonal distortions.

\begin{figure}
\begin{center}
\includegraphics[trim=1.5cm 4.5cm 1.5cm 5.5cm, clip=true, width=8.5cm]{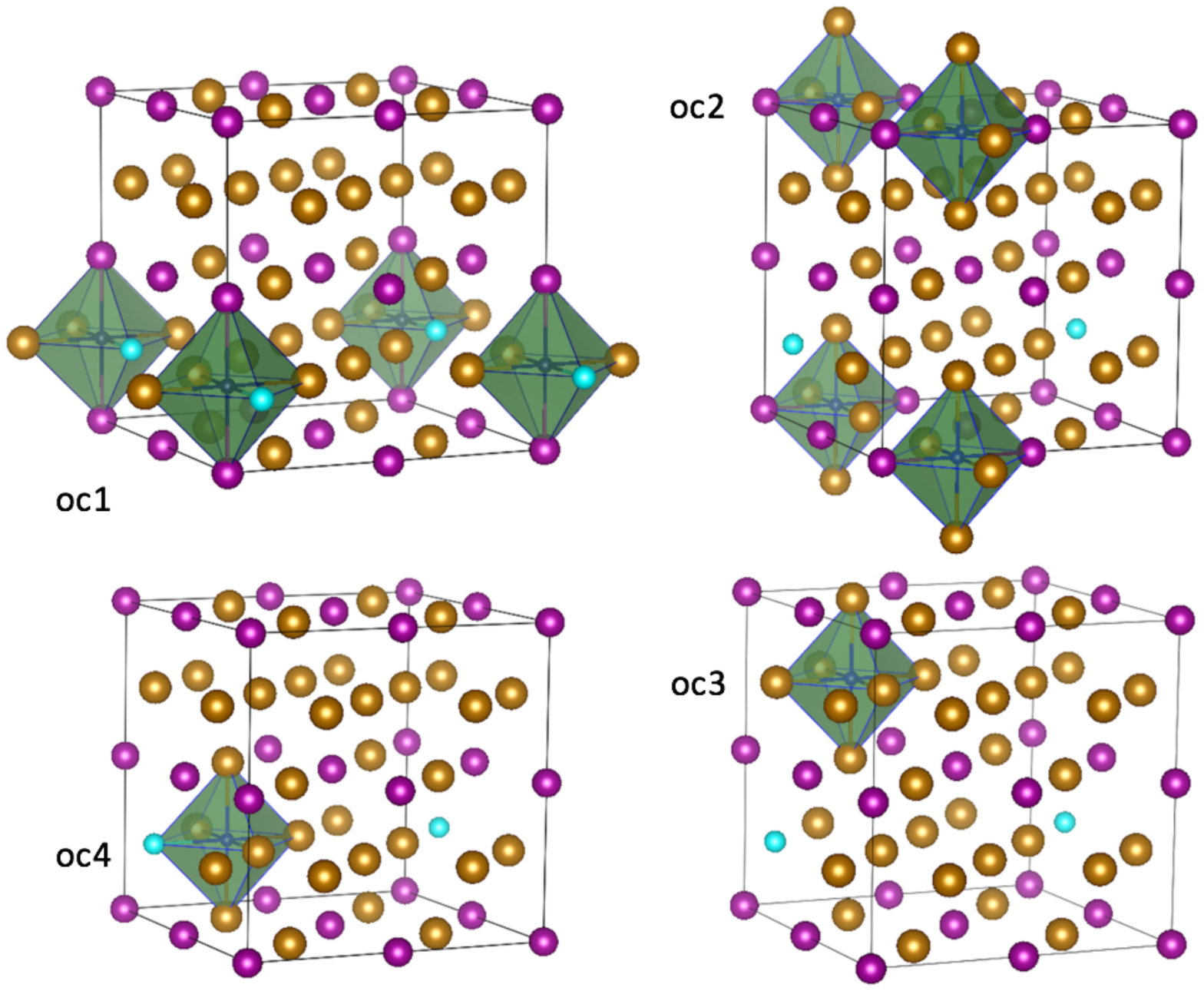}
\caption{The Fe$_{23}$Mn$_8$Al austenitic supercell. Fe atoms are yellow, Mn atoms are purple and blue marks Al. 
The four different octahedral sites we consider are highlighted by polyhedra. 
The hydrogen atom can be located in direct neighbourhood of 3 Fe, 2 Mn and 1 Al atoms (`oc1'), 4 Fe and 2 Mn atoms (`oc2'), 6 Fe atoms (`oc3') or 5 Fe and 1 Al atom (`oc4'). 
Not shown here are the tetrahedral sites. We note that the oc2 configuration has the same next neighbour configuration as our reference octahedral site in the Fe$_{24}$Mn$_8$ system, 4 Fe and 2 Mn atoms, but differs by the substitution of one Fe by Al. The oc1 case corresponds to the substitution of one Fe atom at the reference site with an Al atom, changing the next neighbourhood configuration as 4 Fe 2 Mn $\to$ 3 Fe 2 Mn 1 Al. 
}
\label{octaSites}
\end{center}
\end{figure}

The central quantity required to determine the equilibrium hydrogen concentrations is the solution energy, which is defined as
\be
E^\sigma = E_{MH}[\sigma] - E_M - \frac{1}{2} E_{H_2}. \label{eq1}
\ee
Here, $E_{MH}[\sigma]$ is the energy of the metal-hydrogen system, where $\sigma$ denotes the occupied interstitial site of the hydrogen, as illustrated in Fig.~\ref{octaSites}.
$E_M$ is the energy of the purely metallic alloy system, and $E_{H_2}/2$ is the reference potential for half an isolated H$_2$ molecule.
We note that for vanishing external stress the solution energy and enthalpy coincide.
This energy difference $E^\sigma$ will in the following be computed using {\em ab initio} techniques.
It serves as an input for the calculation of the equilibrium hydrogen concentration (or occupation probability for the different interstices), which is in the dilute limit of non-interacting hydrogen atoms given by Sievert's law \cite{sieverts1929absorption}, 
\be
\label{Sievert}
c_H \sim p_{H_2}^{1/2} e^{-E^\sigma/k_B T},
\ee
where $p_{H_2}$ is the partial pressure of the surrounding H$_2$ gas.
We have to distinguish this pressure from a mechanical pressure $P$ acting on the solid solution phase, which can be independently controlled by mechanical constraints.
In the following, we will focus on the mechanical part and not further discuss the gas pressure $p_{H_2}$.
The value of the (suppressed) proportionality constant in the above expression (\ref{Sievert}) depends on the choice of a reference pressure and the precise definition of the hydrogen concentration.

%
%
%
%

\subsubsection{Stress free systems}

We calculate the energy contributions $E_{MH}$ and $E_M$ for the supercells shown in Figs.~\ref{fig10} and \ref{octaSites}, using full ionic relaxation and vanishing external (mechanical) pressure, $P=0$.
In this case the solution energies are identical to the solution enthalpies.
The resulting values are given in Table \ref{solEntsHydroTable}.
 \begin{table}
\begin{center}
 \caption{The solution enthalpies of hydrogen at the four investigated octahedral sites and in the Fe-Mn reference system.
  The second row describes the nearest neighbor (n.n.)~chemical environment of the hydrogen atom in the octahedral position, where e.g.~4-2-0 denotes that it is surrounded by 4 Fe, 2 Mn and no Al atoms.}
  \begin{tabular}{ || l || l || l || l || l || l ||}
    \hline
    $\sigma$ & Fe-Mn & oc1 & oc2 & oc3 & oc4 \\ \hline
    n.n. (Fe-Mn-Al) & 4-2-0 & 3-2-1 & 4-2-0 & 5-0-1 & 6-0-0 \\ \hline
    $E^{\sigma}(P=0)$ (eV) & 0.054 & 0.140 & -0.010 & 0.130 & 0.073    \\ \hline 
  \end{tabular}
   \label{solEntsHydroTable}
\end{center}
   \end{table}

For the Fe-Mn reference system we only consider the energetically most favorable configuration shown in Fig.~\ref{fig10}, using a Fe$_{24}$Mn$_8$ supercell, where the hydrogen atom in the octahedral position is surrounded by 4 iron and 2 manganese atoms.
Such a site exists 24 times in the supercell.
The other type of octahedral sites, where the H atom is surrounded by 6 Fe atoms, exists 8 times, but has a higher solution energy when the system is deformed.

This energy is compared to different atomic arrangements for a Fe$_{23}$Mn$_8$Al system, where one iron atom is replaced by aluminum.
For this configuration, there are four  chemically inequivalent environments for the H atom, as depicted in Fig.~\ref{octaSites}.
For the configuration denoted by `oc1' the nearest neighbors of the the H atom are 3 Fe, 2 Mn and one Al atom, which exists four times in the supercell.
The configuration `oc2' has the neighborhood 4 Fe, 2 Mn, 0 Al, with a multiplicity of 20.
`oc3' has the environment 6 Fe, 0 Mn, 0 Al, and it exists six times.
Finally, `oc4' has 5 Fe, 0 Mn, 1 Al neighbors and exists twice. Apart from the chemical environment, also the local geometric arrangement influences the solution energies. In the case of the `oc2' configuration, which is of specific interest due to its energetic preferrability, we find the most attractive local arrangement in 9 of the 20 sites. The resulting 11 sites are less attractive and split into three subgroups of local geometric arrangements. 
The resulting solution energies for the most attractive sites for each chemical environment are listed in Table \ref{solEntsHydroTable}.

By comparison to the reference system Fe$_{24}$Mn$_8$, we see that there are two types of interstices with identical next neighbour configuration there and in Fe$_{23}$Mn$_8$Al. 
The oc2 site is energetically more favourable by about 60 meV than the reference site with 4 Fe and 2 Mn atoms as nearest neighbor. 
This is attributed to the nonlocal effect of the substitutional Al atom. The oc3 site is about 70 meV less attractive than the reference site with 6 Fe atoms as nearest neighbor, which has nearly the same energy as the reference site with 4 Fe and 2 Mn atoms in the immediate neighborhood. 
This discrepancy shows that the solution energy is lowest in a manganse rich environment, which is in agreement with the predictions in \cite{ismer2010ab, von-Appen:2014aa}. It also exhibits that in the Fe$_{23}$Mn$_8$Al system, the volumetric effect of the Mn atoms is limited to the direct vicinity of the Mn atoms.

The general observation is therefore, that Al atoms as next neighbour make octahedral sites energetically less favorable.
We can therefore conclude that for low temperatures and hydrogen concentrations the sites close to aluminum atoms will be not be populated.
In the room temperature regime with $k_B T\approx 25\,\mathrm{meV}$ the occupation probability ratio of the energetically most attractive oc2 configuration of the Fe-Mn-Al system in comparison to the Fe-Mn reference system is therefore according to Eq.~(\ref{Sievert}) given by
\begin{equation}
\frac{c_{oc2}}{c_{ref}} = r_\mathrm{oc2/ref} \exp \left(\frac{E^{oc2} - E^{ref}}{k_B T}\right)\approx 3.3,
\end{equation}
where the factor $r_\mathrm{oc2/ref}$ takes into account the multiplicity of the amount of reference sites and relevant `oc2' sites.


Since for chemical trends only differences of solution energies are important, not the absolute values, the contribution of quantum mechanical zero point vibrations is neglected.
In fact, they turn out to be only about 0.02 eV as upper limit e.g.~for Fe$_{1-x}$Mn$_x$ at $x=0,1$ as reported in \cite{ismer2010ab}. 
Similar to the arguments in \cite{ismer2010ab} we can therefore ignore this contribution.

In the calculations we have ignored magnetic effects.
The antiferromagnetic double layer ordering has been shown to be the energetically most stable collinear configuration \cite{sjostedt2002noncollinear} and a good approximation to the true ground state in $\gamma$ Fe. 
However, we recall that two important effects of magnetism, namely the repulsion of the hydrogen atoms by the magnetic moment and the increased interstitial volume due to the larger lattice constant with increasing magnetic ordering -- from nonmagnetic, via antiferromagnetic to ferromagnetic -- are large by magnitude, but they mainly compensate each other. 
We refer to Refs.~\cite{ismer2010ab,song2014interaction} for details about the nonmagnetic approximation.

\subsubsection{Isotropic deformation}

In contrast to the calculations above, where we used a fixed pressure $P=0$, we now fix the volume of the supercell.
This means, that we evaluate the energies $E_{MH}[\sigma]$ and $E_M$ both at the same lattice constant.
Specifically, for the equilibrium lattice constant of the system with hydrogen, the purely metallic system is therefore under tension and has a higher energy due to this elastic effects, compared to the results in Table \ref{solEntsHydroTable}, where both energies were calculated at their individual equilibrium volumes.
Consequently, the solution energies are lower for the present fixed volume boundary conditions.

\begin{figure}
\begin{center}
\includegraphics[width=9cm]{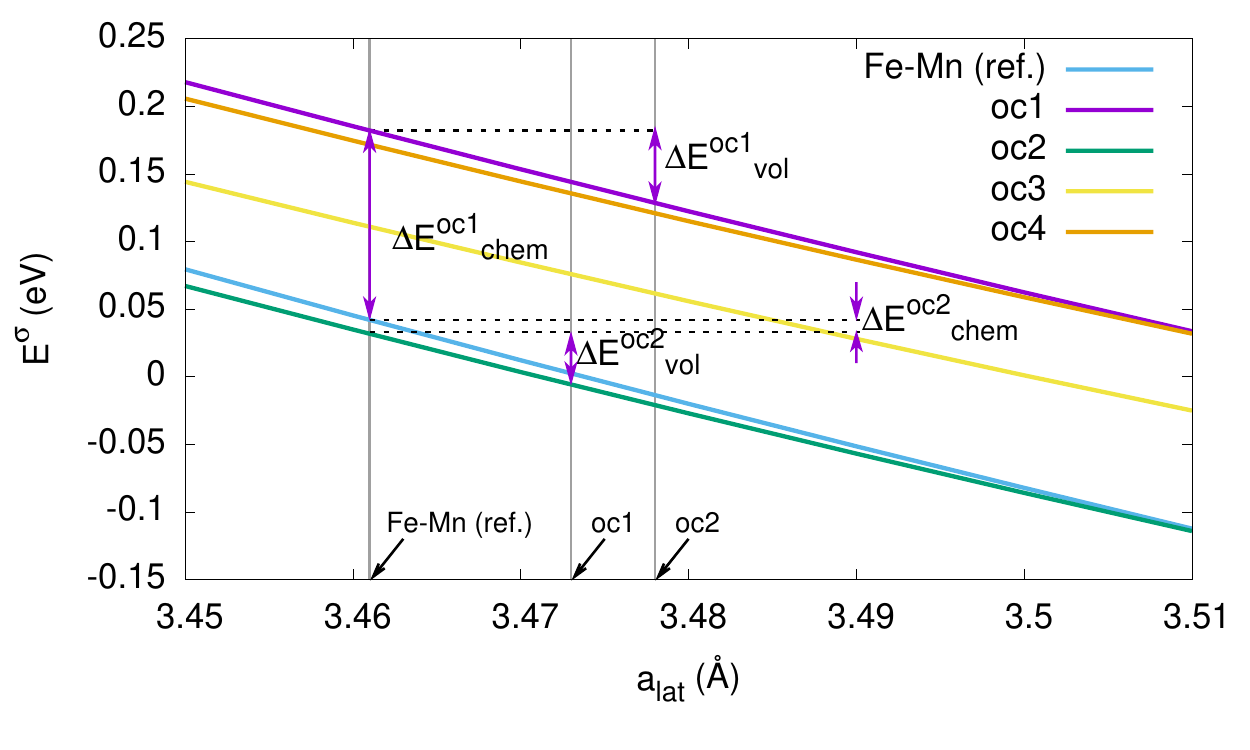}
\caption{Hydrogen solution energies for the different octahedral interstitial sites in Fe$_{24}$Mn$_8$ and Fe$_{24}$Mn$_8$Al versus lattice constant $a_\mathrm{lat}$.
The different curves demonstrate the effect of aluminum on the solution energy. 
The latter is split into a chemical and a volumetric contribution, see main text.
For all configurations the variation of the solution energy with the lattice constant is similar, as well as the volumetric contributions, whereas the chemical contributions differ significantly, indicating the strong preference of H to avoid an aluminum rich nearest neighbour environment.
The vertical lines show the equilibrium lattice constants of the different configurations.
}
\label{solEntHydro42_420_321}
\end{center}
\end{figure}

For isotropic deformations we strain the supercell equally in all three principal directions, which can be described by the diagonal transformation tensor
\be
\label{tensor2}
\underline{\underline{\delta}}_{hydro} = (1+\delta) \cdot \underline{\underline{1}}.
\ee
The resulting solution energies are shown in Fig.~\ref{solEntHydro42_420_321} for the different configurations in Figs.~\ref{fig10} and \ref{octaSites}.
The first observation is that to good accuracy all energies follow a linear behavior in the interval of investigated lattice constants $a_\mathrm{lat}$ with almost the same slope, hence only having a parallel offset.
This is in agreement with the generic picture that hydrogen mainly leads to an isotropic expansion of the stress free system by an amount $\Delta V$, which differs only weakly for different metals \cite{Fukai:2005aa}.
Hence the coupling to an isotropic strain leads to an energy term proportional to $P\Delta V$.
With the pressure $P$ being proportional to the deviation of the lattice constant from its equilibrium value, we therefore expect a linear dependence of the solution energy on the lattice constant.

We can further support this picture by a separation of the solution energy in a volumetric (elastic) and a chemical contribution, as indicated by the vertical and horizontal arrows in Fig.~\ref{solEntHydro42_420_321} respectively.
For the volumetric contribution we consider the hydrogen solution energy difference of the chemically identical system, but evaluated once at its equilibrium lattice constant and once at the lattice constant of the Al free reference system.
Hence, this energy difference contains the release of mechanical work due to the adjustment to the appropriate lattice constant. 
In contrast, the chemical contribution is calculated from the solution energy difference of two chemically different systems, namely the Fe-Mn-Al and the Fe-Mn systems, but this time evaluated at the same lattice constant.
Hence, this energy contribution expresses the energy difference for insertion of a single hydrogen atom into the same rigid atomic structure, but with different chemical environment due to the exchange of Fe by Al.

As can be seen from Fig.~\ref{solEntHydro42_420_321} exemplarily for the configurations oc1 and oc2 is that volumetric contribution is basically identical, in agreement with the above argument, but the chemical contributions differ significantly.
This difference is in line with the stress free results in Table \ref{solEntsHydroTable}. Therefore, also under isotropic strain we find that hydrogen prefers a direct environment which does not contain aluminum.



\subsubsection{Tetragonal deformation} 
 
 Again, the starting point for the consideration is the system with a Fe$_{24}$Mn$_8$ supercell with cubic symmetry.
 By the substitution of one of the Fe atoms by Al, the supercell experiences in equilibrium a shape relaxation with a slight tetragonal distortion.
 On a larger scale the Al atoms will be randomly distributed, and therefore the effects from the different specific arrangement will average out.
 As a result, the supercells will mainly preserve their cubic shape, which we take as basis for additional external tetragonal distortions. 
 
\begin{figure}
\begin{center}
\includegraphics[width=8.5cm]{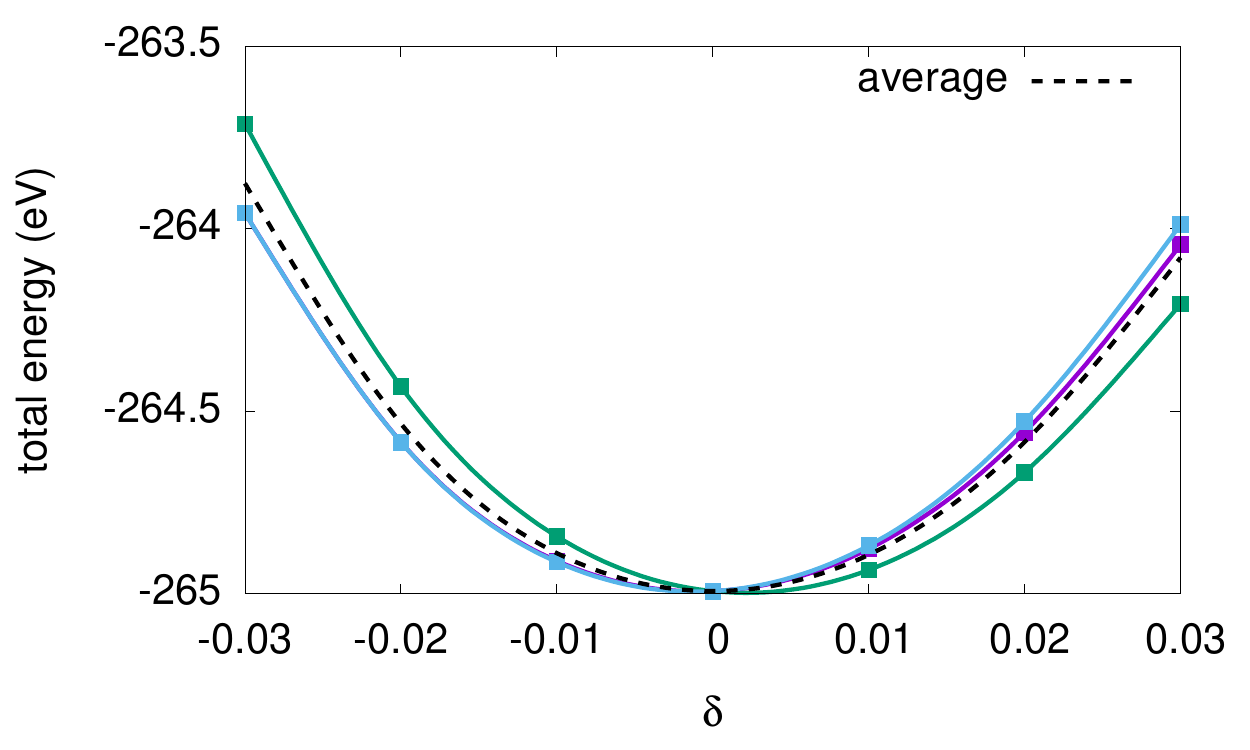}
\caption{The total energy of the Fe$_{23}$Mn$_8$Al system with hydrogen at the oc1 site. 
The coloured curves show the dependence of the energy on the orientation of the strain relative to the preferred elongation axis of the system. 
The dashed black curve shows the averaged energy.
}
\label{strainEnergyAveraging}
\end{center}
\end{figure}

In contrast to the isotropic deformations in the previous section we consider here a case where the distortion is volume-conserving, to clearly demonstrate the different behavior. 
For that, the supercell undergoes a transformation as described by the tensor
%
\begin{equation}
\underline{\underline{\delta}}_{tetra} = 
\begin{pmatrix}\label{tensor1}
1+\delta & 0 & 0  \\
0 & 1+\delta & 0  \\
0 & 0 & (1+\delta)^{-2}
\end{pmatrix},
\end{equation}
which is volume conserving up to third order in the tetragonal strain $\delta$. 
The supercell is stretched in two principal directions and compressed in the other.
The aforementioned dependence of the small intrinsic tetragonal distortion is equivalent to the application of different directions of the compression.
The resulting energy curves as function of the tetragonal strain $\delta$ are shown in Fig.~\ref{strainEnergyAveraging} for the Fe$_{23}$Mn$_8$AlH supercell in configuration oc1.
It shows the characteristic quadratic dependence of the elastic energy as function of strain when deformed from the cubic reference state.
The dashed curve is the arithmetic means of the cases with the three different orientations of the principal axes and reflects the averaged elastic response for a larger system with random distributions.
Such an averaging will therefore also be employed for the following calculations, while we still distinguish between the different configurations shown in Fig.~\ref{octaSites}.
 


\begin{figure}
\begin{center}
\includegraphics[width=8.5cm]{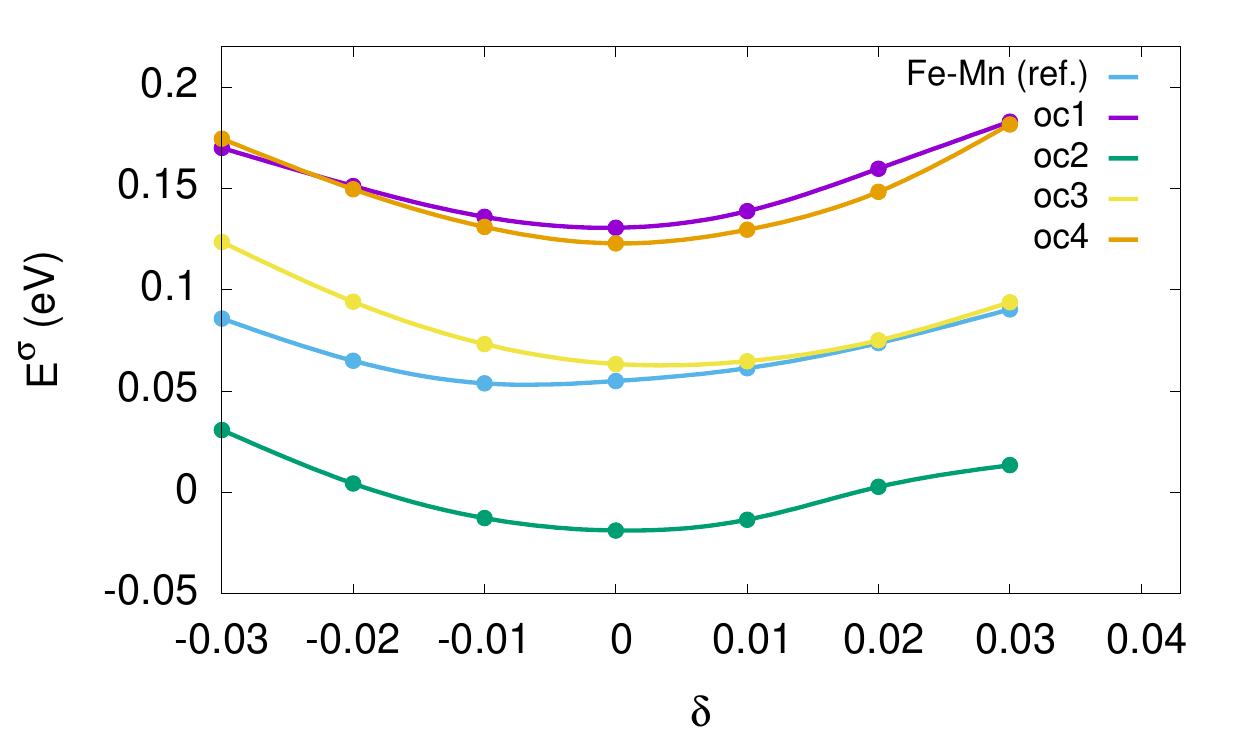}
\caption{Hydrogen solution energy for the octahedral interstitial sites as function of the applied tetragonal distortion.
The effect of the strain is much smaller than the chemical effect due to the substitution of Fe by Al.}
\label{solEnttetra}
\end{center}
\end{figure}

Similar to the isotropic case we computed the hydrogen solution energy as function of the tetragonal distortion, and the results are shown in Fig.~\ref{solEnttetra}.
Obviously, the behavior is very different than for the isotropic deformation in Fig.~\ref{solEntHydro42_420_321}.
For that we note that the considered strain interval $\Delta\delta=0.06$ in the present plot is much larger than in the previous one, where he had $\Delta\delta\approx 0.017$.
Still, the energetic variation is much smaller here, despite the larger strain interval.
Together with the shape of the curves we can therefore conclude that strain effects to the solution energy are a higher order correction here.
This is in agreement with the argument above that the interstitial hydrogen mainly leads to an isotropic expansion of the system, and therefore it only weakly interacts with a volume conserving external strain.

As a result, the chemical influence of the Al atom is clearly dominant compared to the contributions from the tetragonal deformation. 
The hydrogen sites in the vicinity of the Al atom, oc1 and oc4, exhibit solution energies which are not distinguishable within the accuracy limits of our approach. 
The oc3 site is comparably attractive as the reference site, while the oc2 site exhibits the shift of about -60 meV relative to the reference system basically over the full range of tetragonal deformations.

\subsection{Hydrogen diffusion}

In this section we intend to obtain information on the hydrogen diffusion coefficient for different strain states and to understand the influence of aluminum.
For that, we consider diffusion of hydrogen in the Fe-Mn(-Al) matrix via trap site and Einstein diffusion. 
To discuss the influence of aluminum induced trapping we apply Oriani's theory \cite{oriani1970diffusion}, which is applicable when the trap sites and the regular lattice sites for the diffusing species are in local equilibrium. The term local equilibrium means that the kinetics of the equilibration of the regular sites and trap sites are described by sufficiently separated timescales. 
Then, important parameters are the solution energies and the trap densities.
This allows to draw conclusions on the relative change of the diffusion coefficients depending on composition and strain.

For Einstein diffusion, we obtain estimates for the changes of the diffusion coefficients based on the diffusion barriers of the involved transition paths. 
We employ climbing image nudged elastic band calculations (CINEB) \cite{henkelman2000climbing,henkelman2000improved}, which search the saddle point of the reaction path, i.e.~the barrier between two connected sites. 

In the spirit of the above discussion we use initial octahedral and final tetrahedral states of the reaction path with cubic cell symmetry and averaged lattice constants, and therefore can refrain from using solid state nudged elastic band calculations \cite{sheppard2012generalized}.
The energy difference due to the averaged lattice constant -- e.g. $a_\mathrm{lat}=3.502\,\angstrom$ for Mn$_{32}$H with hydrogen in an octahedral position and $a_\mathrm{lat}=3.506\,\angstrom$ for occupation of a tetrahedral site, leading to an averaged value $\bar{a}_\mathrm{lat}=3.504\,\angstrom$ --  is in the range of 5 meV. 
We used Mn$_{32}$H to test the validity of this approximation against available data on transition points and found good agreement \cite{ismer2010ab}. 
We also have checked the convergence of the CINEB calculations with respect to the number of images which initially discretise the transition path. 
%
 

\subsubsection{Stress free systems}

The oc2 sites are traps with a strain dependent binding energy.
In absence of any deformation they have a depth of 60 meV, relative to the sites in the reference system Fe$_{24}$Mn$_8$.  
There are 20 possible hydrogen oc2 sites which are preferential and we assume that at most half of them are accessible due to the indirect hydrogen repulsion when placed in adjacent sites. 
This leads to the estimate of the trap density as $N_t = 1.34 \times 10^{28}\,\mathrm{m}^{-3}$, and the $N_l$ is the density of regular sites, with the ratio $N_t/N_l \approx 0.28$. 

\begin{figure}
\begin{center}
\includegraphics[width=8.5cm]{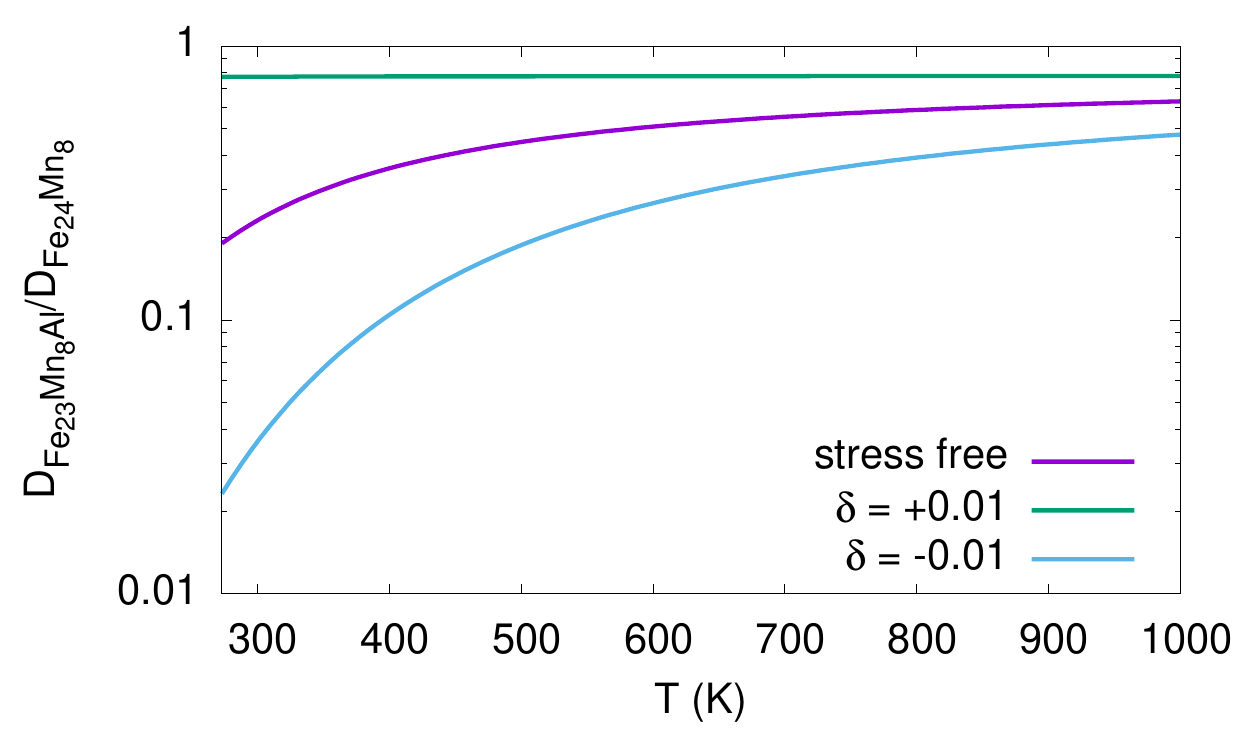}
\caption{Ratio of the diffusion coefficients according to Oriani's theory as described by Eq.~(\ref{orianiRatioDiffusion}). The  isotropic deformation leads to changes of the apparent diffusion coefficient by about one order of magnitude for a strain $\delta=\pm 0.01$ at room temperature. 
In all cases the presence of Al reduces the diffusion coefficient also by one order of magnitude in comparison to the Fe$_24$Mn$_8$ at ambient temperature.
}
\label{orianiHydro}
\end{center}
\end{figure}

The apparent diffusivity relative to the trap-free diffusivity (aluminum free) diffusivity is according to Oriani's theory
\be
\label{orianiRatioDiffusion}
\frac{D_{\mathrm{Fe}_{23}\mathrm{Mn}_8\mathrm{Al}}}{D_{\mathrm{Fe}_{24}\mathrm{Mn}_8}} = \frac{1}{\displaystyle 1+\frac{N_t}{N_l}\exp\left( \frac{E^\mathrm{oc2}-E^\mathrm{ref}}{k_B T}\right)},
\ee
which is plotted in Fig.~\ref{orianiHydro} as function of temperature.
We see that the presence of aluminum suppresses the hydrogen diffusivity in the room temperature range by about one order of magnitude.

\subsubsection{Isotropic deformation}

Before discussing the influence of an isotropic deformation on trap site diffusion, we mention that differences in the bulk moduli between the hydrogen free and hydrogen loaded systems are typically small, see Table \ref{murnaghanTable}. 
\begin{table}
\begin{center}
 \caption{Energies at equilibrium volume, bulk modulus and its derivative with respect to pressure as obtained from Murnaghan fits of the {\em ab initio} data. 
}
  \label{murnaghanTable}
  \begin{tabular}{ || l || l | l | l || }
    \hline
    System & $E_0 (\mathrm{eV})$  & $B_0 (\mathrm{GPa})$  & $\partial B/\partial P$     \\ \hline
    Fe$_{24}$Mn$_8$ & -265.581  & 176.6 & 4.78   \\ \hline 
    Fe$_{24}$Mn$_8$H & -268.922  & 175.6 & 4.70    \\ \hline 
    Fe$_{23}$Mn$_8$Al & -261.717  & 168.5 & 4.847   \\ \hline 
    Fe$_{23}$Mn$_8$AlH (oc1) & -264.992  & 168.3 & 4.763    \\ \hline
     Fe$_{23}$Mn$_8$AlH (oc2) & -265.141  & 167.4 &  4.910    \\ \hline 
  \end{tabular}
\end{center} 
  \end{table}
Thus, equal strains correspond in good approximation to equal pressures.



Fig.~\ref{orianiHydro} shows additionally to the stress free case the ratio of the trap site diffusion coefficient according to the Oriani model for isotropic compression ($\delta=-0.01$) and tension ($\delta=0.01$).
This leads to a change of the diffusion coefficient relative to the case without aluminum by up to one order of magnitude at room temperature, based on the consideration of the oc2 traps.
In general, a tensile strain increases the diffusion coefficient, as the larger interatomic spacing allows for easier jumps of the interstitial hydrogen via a reduction of the energy barriers.


For Einstein diffusion we focus on the configurations oc1, oc2 and the reference case of the supercell without Al, as these are most illustrative to discuss the influence of the Al alloying with respect to deformation effects.

For isotropically strained systems with only one Al atom in the 32 atom supercell, the diffusion coefficient behaves as
\begin{equation}
D \sim n_{tb} a^2 \exp \left(-\frac{E_\mathrm{Diff}}{k_B T} \right),
\end{equation}
and both changes of the exponent and the pre-exponential factor for variations of composition and strain state have to be considered.
Here, $n_{tb}$ is the number of transition bonds to the next sites, $a$ is the distance to these sites, and $E_\mathrm{Diff}$ is an effective diffusion barrier which reflects the dominant transitions contributing to the diffusion process.  

\begin{figure}
\begin{center}
\includegraphics[width=8.5cm]{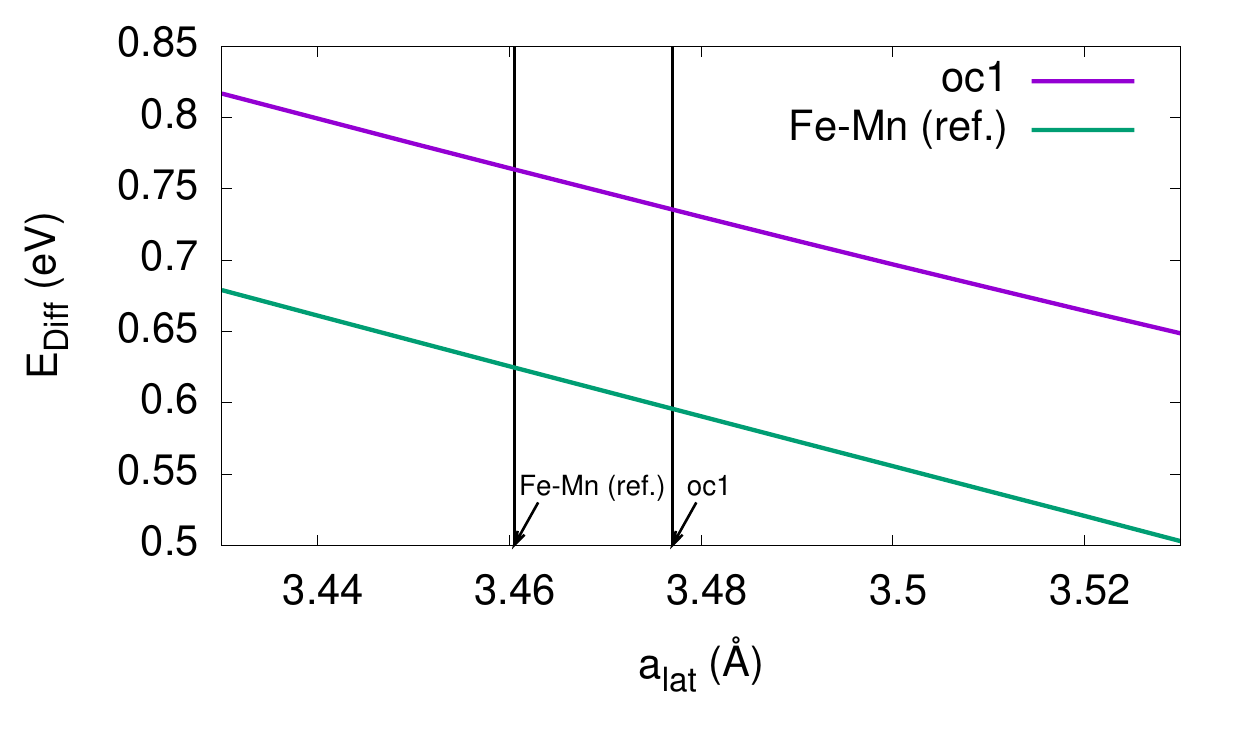}
\caption{The diffusion barriers $E_\mathrm{Diff}$ for the system before Al alloying, see the lower curve, and the site oc1 in vicinity of the Al atom, see the upper curve. The grey bars mark from left to right the optimal volume of the reference site and oc1. We recall that at zero pressure, the shift in barrier height due to substituting an Fe atom in the reference system by an Al atom is about 126 meV.}
\label{barriersHydro}
\end{center}
\end{figure}

In Fig.~\ref{barriersHydro} we show the barriers, i.e.~the energy difference $E_\mathrm{Diff}$ between the transition and starting point, as function of lattice constant.
The starting point is a hydrogen atom in position oc1, which then crosses a tetrahedral site close to the Al atom, with one Fe, two Mn and one Al atom as next neighbours. This initial site is relevant when local accumulations of Al yield a low density of available more attractive sites. 
The barrier of this transition close to the Al atom is increased by about 126 meV, which means that this transition will play a negligible role when other transitions are accessible. Under expansion, the system yields a substantial decrease of the barrier height. 

The density of high-barrier transitions in the system with an Al atomic concentration of $1/32$ is still low, and therefore one cannot assume that they efficiently reduce the number of transition paths from the considered site in the spirit of shutting down percolation paths.
Therefore, this leads only to a weak decrease of the pre-exponential factor $n_{tb}$.
However, when the number of high-barrier transition paths is increased, i.e.~when the atomic concentration of Al is increased to 2/32 or 3/32, the probability that only the high barrier transitions along sites of type oc1 are accessible from a considered site increases.
Under these circumstances the energy barriers related to jump processes starting at oc1 type positions become dominant, which leads to a change of the relevant exponential factor.
It leads to a reduction of the diffusion coefficient relative to the aluminum free system by a factor
\begin{equation}
\exp\left( -\frac{E_\mathrm{Diff}^\mathrm{oc1} - E_\mathrm{Diff}^\mathrm{ref}}{k_B T} \right),
\end{equation}
which can reach $\sim 10^{-3}$ at room temperature.

\subsubsection{Tetragonal strain}

\begin{figure}
\begin{center}
\includegraphics[width=8.5cm]{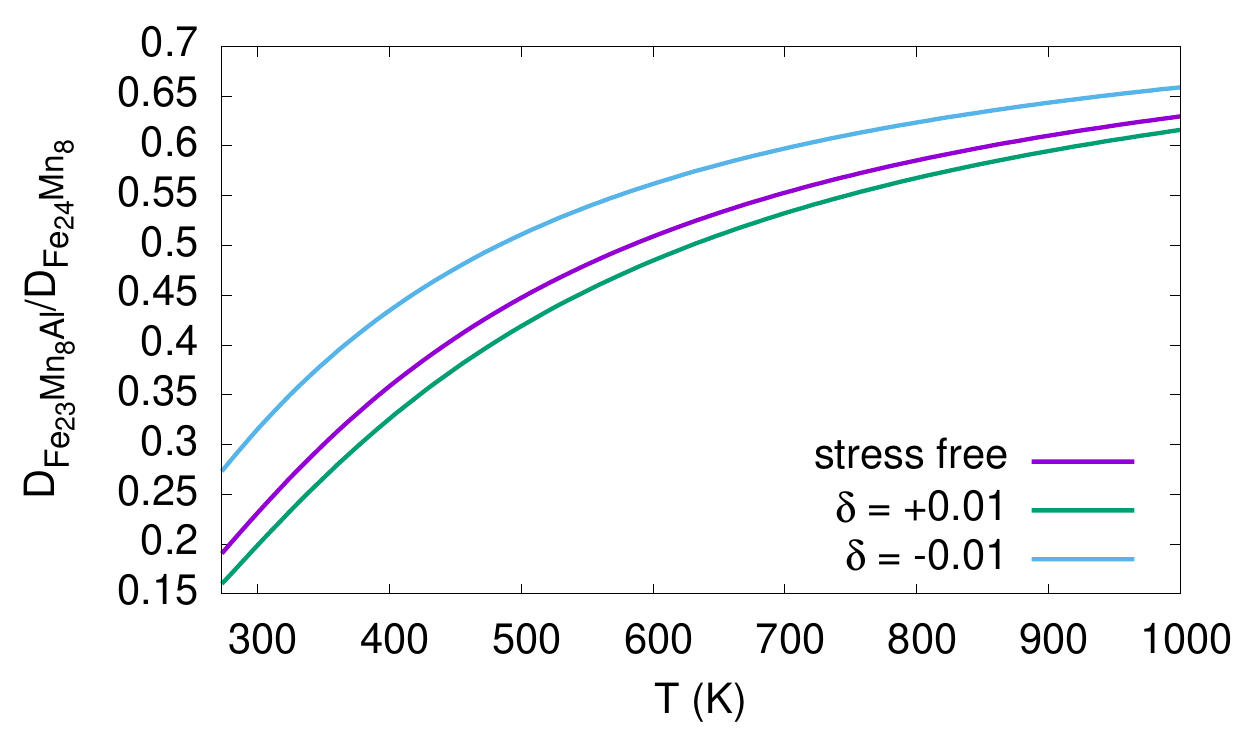}
\caption{Ratio of diffusion coefficients corresponding to Oriani's theory for tetragonal strain. We note that the tetragonal deformation has a comparably weak effect on the apparent diffusivity. The ratio of the diffusion coefficients in the deformed systems varies by less than one order of magnitude.}
\label{orianitetra}
\end{center}
\end{figure}

We can proceed in a similar way for systems under tetragonal strain.
We distinguish between the three different possible strains relative to the elastically stiff axis, as they all appear in a random distribution of Al in an Fe-Mn-Al alloy. 
As expected, the effect of the deformation on the trap site diffusion is weak, see Fig.~\ref{orianitetra}.
At room temperature, the ratio of the apparent diffusion coefficients is changed by less than order of magnitude. 

\begin{figure}
\begin{center}
\includegraphics[width=8.5cm]{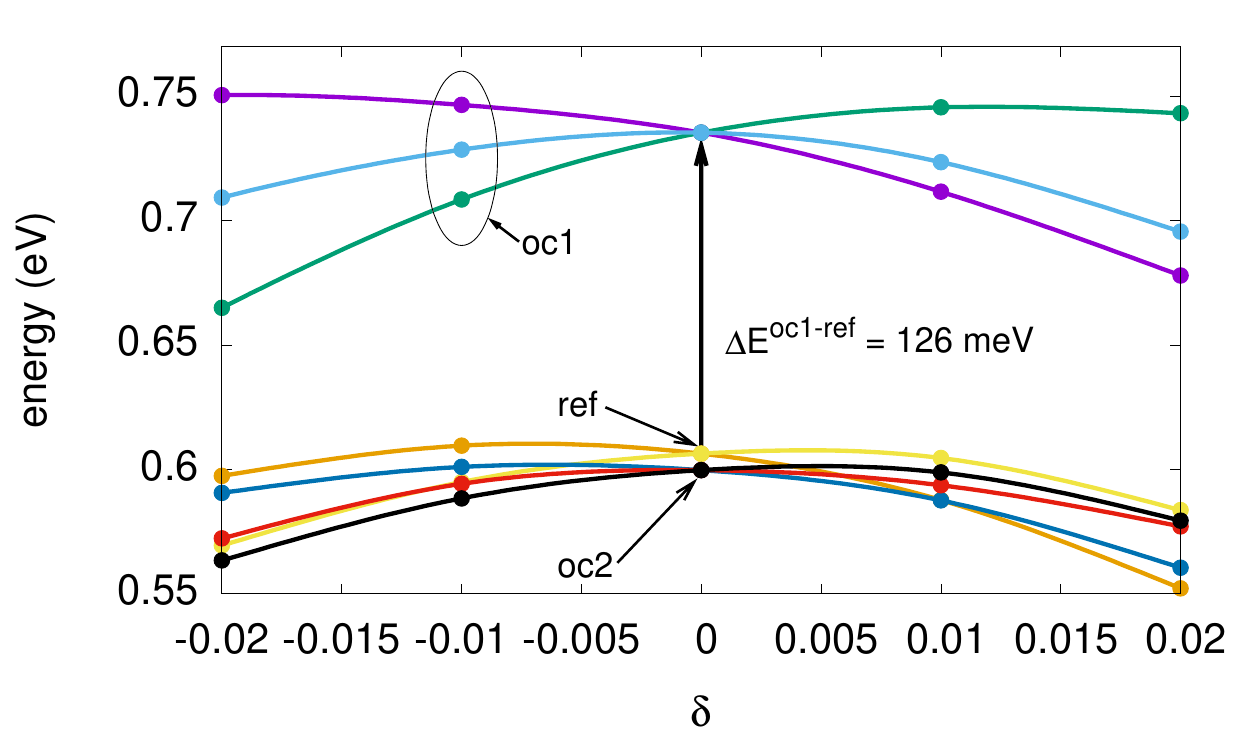}
\caption{The diffusion barriers $E_\mathrm{Diff}$ for the system before Al alloying, see the two middle curves in yellow and orange (ref), the diffusion barrier for the alloyed system from the octahedral site oc1 in direct vicinity of an Al atom to a tetrahedral site with 1 neighbouring Fe, 2 neighbouring Mn and 1 neighbouring Al atoms (three upper curves oc1), and the diffusion barriers for the alloyed system from the octahedral site oc2 distant to an Al atom to a tetrahedral site with 2 neighbouring Fe and 2 neighbouring Mn atoms (three lower curves oc2). Note that the shift in the diffusion barrier for the sites in the vicinity of the Al atom is about 126 meV. We recognise that the orientation of the tetragonal strain relative to the stiff axis of the system has appreciable influence especially for sites in the neighbourhood of the Al atom.
}
\label{einsteinDiffusion}
\end{center}
\end{figure}

The picture is different when we consider the effect of the deformation on Einstein diffusion processes. 
As shown in Fig.~\ref{einsteinDiffusion}, variations in the relative orientation of the preferred elongation axis can result in a shift of the diffusion barriers by two orders of magnitude at low temperatures. 
The chemical shift at zero pressure is larger, yielding a possible decrease of the diffusion coefficient in the vicinity of Al atoms up to three orders of magnitude. Again, the chemical contribution to the solution energy is dominant in the non-stressed state.

\section{\RNum{4}. Summary \& conclusions}

In the present paper we have discussed the influence of Al alloying and deformations on hydrogen absorption and diffusion in high manganese steels, represented by Fe$_{24}$Mn$_8$ and Fe$_{23}$Mn$_8$Al systems. 
In the considered configurations, the substitution with Al leads to two counteracting effects on the hydrogen dissolved on interstitial sites in the matrix. 
On the one hand, at low Al concentrations of about 3at\%, the number of trap sites, which increase the hydrogen solubility and decrease the diffusivity, is large. 
On the other hand, at larger Al concentrations an effective blocking of paths for the hydrogen diffusion could lead to a drastic reduction of the diffusivity, caused by the high barrier of the transition path between octahedral and tetrahedral sites in the direct vicinity of aluminum atoms. Here we note that the high Mn contents in TWIP steels allow for high Al additions without destabilizing the austenite.

In general, the influence of volumetric changes during isotropic deformations has larger influence on the diffusivity of hydrogen compared to volume preserving tetragonal strains.
While the volumetric deformation typically results in a linear dependence of the solution energy on the volume, the influence of tetragonal strains yield effects of second order. 
This is in agreement with the notion that hydrogen leads to an isotropic expansion of the host alloy.

Spatially inhomogeneous Al distributions lead to an orientation dependent influence of a tetragonal distortion on the diffusion of hydrogen.
High local aluminum concentrations have the potential to reduce the diffusion coefficient locally by up to three orders of magnitude in the room temperature regime.
For a careful investigation of this effect, large scale Monte Carlo simulations with an emphasis on percolation effects may shed further light on this issue.


\section{Acknowledgements}

This work has been supported by the Collaborative Research Center (SFB) 761 {\em Steel ab initio} of the German Research Foundation (DFG).
The authors gratefully acknowledge the computing time granted on the supercomputer JURECA at the J\"ulich Supercomputing Centre (JSC).
Discussions with Tilmann Hickel are highly appreciated by the authors.

\bibliographystyle{ieeetr}

\bibliography{references.bib}%

%
%
%
%

\end{document}